\documentclass[journal]{IEEEtran}
\usepackage{amsmath,lipsum}
\usepackage{rotating}
\usepackage{amssymb}
\usepackage{graphicx}
\usepackage{makecell}
\usepackage{caption}
\usepackage{subfig}
\usepackage{relsize}
\usepackage[usenames, dvipsnames]{color}
\usepackage{booktabs}% http://ctan.org/pkg/booktabs

\usepackage{enumitem}
\usepackage{booktabs}
\usepackage{mathtools}
\usepackage[english]{babel}
\usepackage{multirow}
\usepackage{lineno}
\usepackage[colorlinks = true,
            linkcolor = blue,
            urlcolor  = blue,
            citecolor = red,
            anchorcolor = blue]{hyperref}

\usepackage{amsmath}
\usepackage{algpseudocode}
\usepackage{algorithm}
\algnewcommand\algorithmicforeach{\textbf{for each}}
\algdef{S}[FOR]{ForEach}[1]{\algorithmicforeach\ #1\ \algorithmicdo}
\algnewcommand\algorithmicinput{\textbf{Input:}}
\algnewcommand\algorithmicoutput{\textbf{Output:}}
\algnewcommand\Input{\item[\algorithmicinput]}%
\algnewcommand\Output{\item[\algorithmicoutput]}%

\ifCLASSINFOpdf
\else
\fi
\begin{document}
\title{Single-modal and Multi-modal False Arrhythmia Alarm Reduction using Attention-based Convolutional and Recurrent Neural Networks 
}

\author{\IEEEauthorblockN{Sajad Mousavi, Atiyeh Fotoohinasab and Fatemeh Afghah}\\
\IEEEauthorblockA{\textit{School of Informatics, Computing and Cyber Systems}, 
{Northern Arizona  University}, Flagstaff, USA \\ \{SajadMousavi, af2329, Fatemeh.Afghah\}@nau.edu}\\ 
% \vspace{+0.2cm}
% \IEEEauthorblockN{U. Rajendra Acharya}\\
% \IEEEauthorblockA{\textit{Department of Electronics and Computer Engineering, Ngee Ann Polytechnic, Singapore} \\
% \textit{Department of Biomedical Engineering, School of Science and Technology, Singapore University of Social Sciences, Singapore}\\
% \textit{Department of Biomedical Engineering, Faculty of Engineering, University of Malaya, Malaysia} \\
% aru@np.edu.sg} 
}


\maketitle

% As a general rule, do not put math, special symbols or citati% in the abstract or keywords.
\begin{abstract}
This study proposes a deep learning model that effectively suppresses the false alarms in the intensive care units (ICUs) without ignoring the true alarms using single- and multi- modal biosignals. Most of the current work in the literature are either rule-based methods, requiring prior knowledge of arrhythmia analysis to build rules, or classical machine learning approaches, depending on hand-engineered features. In this work, we apply convolutional neural networks to automatically extract time-invariant features, an attention mechanism to put more emphasis on the important regions of the input segmented signal(s) that are more likely to contribute to an alarm, and long short-term memory units to capture the temporal information presented in the signal segments. We trained our method efficiently using a two-step training algorithm (i.e., pre-training and fine-tuning the proposed network) on the dataset provided by the PhysioNet computing in cardiology challenge 2015.
The evaluation results demonstrate that the proposed method obtains better results compared to other existing algorithms for the false alarm reduction task in ICUs.
The proposed method achieves a sensitivity of 93.88\% and a specificity of 92.05\% for the alarm classification, considering three different signals. In addition, our experiments for 5 separate alarm types leads significant results, where we just consider a single-lead ECG (e.g., a sensitivity of 90.71\%, a specificity of 88.30\%, an AUC of 89.51 for alarm type of Ventricular Tachycardia arrhythmia)
 \footnote{This material is based upon work supported by the National Science Foundation under Grant Number 1657260. Research reported in this publication was supported by the National Institute On Minority Health And Health Disparities of the National Institutes of Health under Award Number U54MD012388.}.
\end{abstract}
% \textcolor{magenta}{what are the evaluation metrics, how much better? what are the specific scenarios that the results are significantly better, any specific alarm, or combination of signals. Be more specific about the results } 


% \fi
\IEEEpeerreviewmaketitle

\section{Introduction}
\label{sec:intro}
\IEEEPARstart{T}{he}
electrocardiogram (ECG) is a biomedical signal that includes information about the electrical activity of heart function and heart conditions over a period of time. Monitoring and interpretation of ECG signals serve the most useful tool for medical staff in ICUs to check the patients’ heart condition such as arrhythmia, ventricular hypertrophy, and myocardial infarction, etc. Cardiac arrhythmias can cause serious and even potentially fatal symptoms if they are not inspected promptly. %Patient monitoring alarms in ICU enables medical staff to be notified from the upcoming life-threatening arrhythmias. 
Although patient-monitoring alarms play an indispensable role in saving patients' lives, the high rate of false alarms not only can be annoying for patients but also may delay the response of medical staff due to making them less sensitive to warnings. Moreover, it can delay patients' recovery by causing sleep deprivation and depressed immune systems. Therefore, suppressing the rate of false alarms in ICUs will improve the quality of patient care and reduce the number of missed true fatal alarms by medical staff. As reported by Aboukhalil et al. \cite{aboukhalil2008reducing} and Drew et al. \cite{drew2014insights}, the rate of false alarms in ICUs reaches as high as almost 90\%. In regard to this concern, the PhysioNet directed the challenge 2015 to reduce the incidence of false arrhythmia alarms in ICUs while the true alarms are not suppressed \cite{PhysioNetFL}.

In order to reduce the rate of false alarms in ICUs, various methods have been proposed. Typically, they can be classified into two general categories: 1) methods based on cardiac rules and 2) machine learning based methods. In the first category, some cardiac rules are defined by experts to detect alarm types. All of the approaches in this category depend primarily on the QRS-complex detection in order to estimate heart rate (HR) and evaluate the signal quality. Ansari et al. \cite{ansari2015multi} adopted several peak detection algorithms to create a robust peak detection algorithm and exploited the information from all three ECG, ABP and PPG signals. Fallet et al. \cite{fallet2015multimodal} used an adaptive frequency tracking algorithm to estimate HR from PPG and ABP signals and an adaptive mathematical morphology approach to estimate HR from the ECG. Also, they exploited the Spectral Purity Index (SPI) to quantify the morphological changes of QRS complexes related to the Ventricular Arrhythmia. Then, they employed a set of rules based on the HR and the SPI to inspect false alarms. Plesinger et al. \cite{plesinger2015false} and Couto et al. \cite{couto2015suppression} applied a set of rules on each alarm types to distinguish between false and true alarms using ECG, ABP and PLETH signals. He et al.\cite{he2015reducing} classified alarms using ECG and ABP signals by following a set of rules related to Signal Quality Index (SQI) and Heart Rate Variability (HRV). However, one challenge with the false alarm detection based on cardiac rules is the need for an expert to determine the rules and the required thresholds. To tackle this, recent studies have exploited machine-learning approaches to detect false alarms.

Machine learning field has many different applications ranging from computer vision, wireless and IoT networks, Unmanned aerial vehicle (UAV) navigation to clinical prediction models \cite{mousavi2017traffic,afghah2018reputation,mousavi2016deep,shamsoshoara2019distributed,shamsoshoara2019solution,mousavi2016learning,mousavi2017applying,shamsoshoara2015enhanced,shamsoshoara2019survey,fotoohinasab2018denoising}. 
In machine learning based methods, a false alarm detection model is trained using some extracted features from the dataset's samples. In \cite{antink2016reducing}, features of interest are extracted from the two-dimensional beat-to-beat correlograms using Fast Fourier Transform (FFT) and principle component analysis (PCA) as well as basic statistical and self-similarity analysis. Then, several machine learning algorithms are evaluated using the extracted features to detect false alarms. In \cite{gajowniczek2019reducing}, a random forest technique is applied to reduce false alarms using different methods of probability and class assignments. Lehman et al. \cite{lehman2018representation} adopted a supervised denoising autoencoder (SDAE) to identify false alarms in Ventricular Tachycardia using features of interest extracted by FFT. Kalidas and Tamil  \cite{kalidas2015enhancing} used a combination of logical and SVM algorithm to classify arrhythmias using ECG and PPG signals. In their work, the features of interest are a set of both time and frequency-domain information. \cite{afghah2015shapley} and \cite{zaeri2018feature} proposed game theoretical approaches in order to extract more discriminative features to reduce the rate of false alarms. 

The performance of the classification methods highly depends on the quality of class discriminating features in terms of on what extent they can capture the main characteristics of the input. Most of the machine learning methods are trained based on hand-crafted features. However, one challenge facing the hand-crafted features is that it depends on a specific dataset, thereby new features may be needed if the dataset changes in terms of size and variety of patients. Although deep learning algorithms have been utilized in medical applications \cite{acharya2019automated,mousavi2019inter,mousavi2018ecgnet},  only a few numbers of studies in the false alarm reduction literature applied deep learning methods and automatic feature extraction \cite{lehman2018representation,hooman2018deep}. In this paper, we propose a deep learning-based approach to reduce the rate of false alarms in ICUs for five life-threatening arrhythmias: Asystole (ASY), Extreme Bradycardia (EBR), Extreme Tachycardia (ETC), Ventricular Tachycardia (VTA), and Ventricular Flutter/Fibrillation (VFB). 
% We leverage deep learning in both feature extraction and classification parts.
The performance of the proposed model is evaluated using the publicly available alarm dataset for ICUs provided by "PhysioNet computing in cardiology challenge 2015". The experimental results show the proposed method can significantly suppress the rate of false alarms in ICU equipment with respect to five mentioned life-threatening arrhythmias without suppressing true alarms. In the following, the main contributions of this work are summarized:

\begin{itemize}
\item
We present a multi-modal model that integrates three main signals of arterial blood pressure (ABP), photoplethysmograph (PPG) and ECG in order to enhance the accuracy of arrhythmia detection and reduce the false alarm rate in ICUs. A multi-modal approach that analyzes a set of independent sources/signals for alarm detection can significantly improve the alarm detection performance. The reason behind this idea is that each independent channel or source of data is inclined to distinct noise and/or artifacts, thereby a hidden pattern in a certain channel caused by noise and/or artifacts can be disclosed by other clean channels.
\item
We develop a network architecture for automatic feature extraction that utilizes a convolutional neural network (CNN) with two consecutive one-dimensional convolutional layers composed of different filter sizes, attention and long short-term memory (LSTM) units, and a classification layer. The CNN part extracts a vector of features from each segment of a single channel, while the attention and LSTM units are trained to identify the most effective parts of the segment in the detection and capture long-range of dependencies between segments of an input signal, respectively. Typically,
some indicators appear in the signals as early as few hours before cardiac events \cite{mozos2015electrocardiographic,abdelghani2016surface,lai2019automated}. Since  considering  the  entire  length  of  the  signals  is  not necessarily  feasible, an attention mechanism along with a memory-based approach can  divide  the signals  into  different  partitions  by  putting  a  higher  weight  on the  most  important  ones  to  save  space/computation  as well as  enhance  the  accuracy. 

% At last, the classification layer determines the label of the input signal. The feature extraction part is trained for the three biosignals separately.
\item
We apply two loss functions of Mean False Error (MFE) and Mean Squared False Error (MSFE) instead of using the common loss function in deep learning algorithms; Mean Squared Error (MSE), to reduce the effect of class unbalanced dataset on degrading the performance. This proposed loss function propagates the training error for a misclassified sample without considering its membership to the major or minor class.
\item
We introduce a two-step training algorithm to train the proposed model effectively so that the effect of the unbalanced class problem is alleviated. In the first step, a model is trained for each of the three signals separately to extract the most distinctive features in regard to each signal. In the second step, the model is trained to generate a label given the three signals.
\end{itemize}

In the next section, we describe the proposed false arrhythmia alarm reduction method. Section \ref{sec:Dataset} provides a description of the dataset used in this study. In Section \ref{sec:expres}, we present the experimental results and compare the performance of the proposed algorithm to other state-of-the-art algorithms, followed by the conclusion in Section \ref{sec:con}. 

%\IEEEpubidadjcol

\section{Methodology}
\label{sec:propsed}
\begin{figure*}[htb]
\centering
  \includegraphics[width=0.75\linewidth,height=0.9\textheight,keepaspectratio]{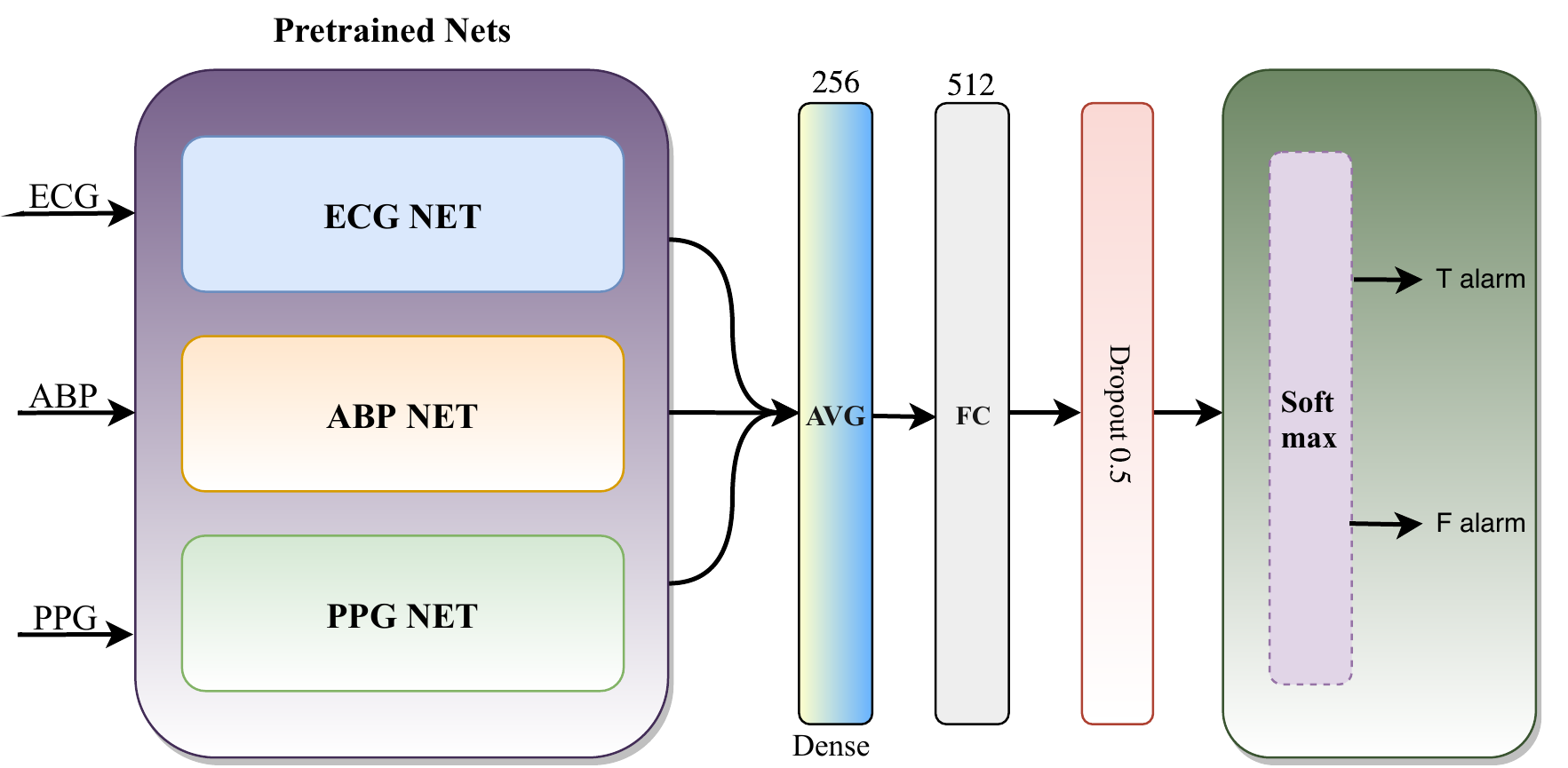}
  \caption{An overview of the network architecture for multi-model false alarm reduction method. AVG: average, FC: fully connected layer.} 
  \label{fig:final-model}
%   \vspace{-15pt}
\end{figure*}

We develop a deep learning model to classify the arrhythmias from the segments of three common physiological signals of ECG, ABP, and PPG signals based on a two-stage approach to further reduce the false alarm rate. In the first part, we develop three pre-trained networks to extract features of interest for the three biosignals separately, followed by a shallow neural network in the second part that uses the extracted features from the pre-trained nets to perform a classification task. At each time step, pre-trained networks extract features of their corresponding input signals, and then, the extracted features are averaged and fed to the fully-connect layer with the size of 256 neurons followed by a dropout block. Finally, a softmax layer is used to determine the probability of the input signal belonging to each class of interest (true or false alarm). Figure \ref{fig:final-model} shows an overall view of the proposed model for reducing false arrhythmia alarms in ICUs. It should be noted that the dropout block is frozen during the testing phase and is just used in the training phase. In the following sections, we describe the details of different parts of the proposed model. 

%Figure \ref{fig:final-model} depicts an overall view of the proposed model for reducing false arrhythmia alarms in ICUs. The model architecture is composed of two main parts: (1) pre-trained networks including ECG, ABP and PPG NETs and (2) a shallow neural network that uses the outputs of pre-trained nets as its input to perform a classification task. In the following sections, we describe a detailed description of our proposed model.

\subsection{Pre-processing}
Prior to feature extraction and classification parts, the ECG, ABP, and PPG signals were subjected to normalization and segmentation steps. For the normalization step, the signals are normalized to a range of 0 to 1. The segmentation part is perfomed using a sliding 200-sample window with an overlap of 25\% for all three signals separately. These segments are fed to their corresponding networks  (i.e., ECG, ABP, and PPG NETs as shown in Figure \ref{fig:final-model}) as the input sequences. It is worth mentioning that the pre-processing process does not include any noise removing and/or filtering steps to remove muscle artifacts and baseline wander.
%The pre-processing process contains these simple following steps. First, we normalized the given signals including ECG, ABP and PPG to a range of between zero and one. Second, the  given  signals were divided into several windows with a size of 200 samples and an overlap of 25\%. These windowed signals (i.e., segments) are used as an input sequence for its corresponding NET (i.e., ECG, ABP, and PPG NETs as shown in Figure \ref{fig:final-model}). It worth motioning that the pre-processing process does not involve any approach to remove noises or any form of filtering for the reduction of baseline wander, muscle artifact, etc. in the signals.
\subsection{The model architecture}
\begin{figure}[htb]
\centering
  \includegraphics[width=\linewidth,height=1\textheight,keepaspectratio]{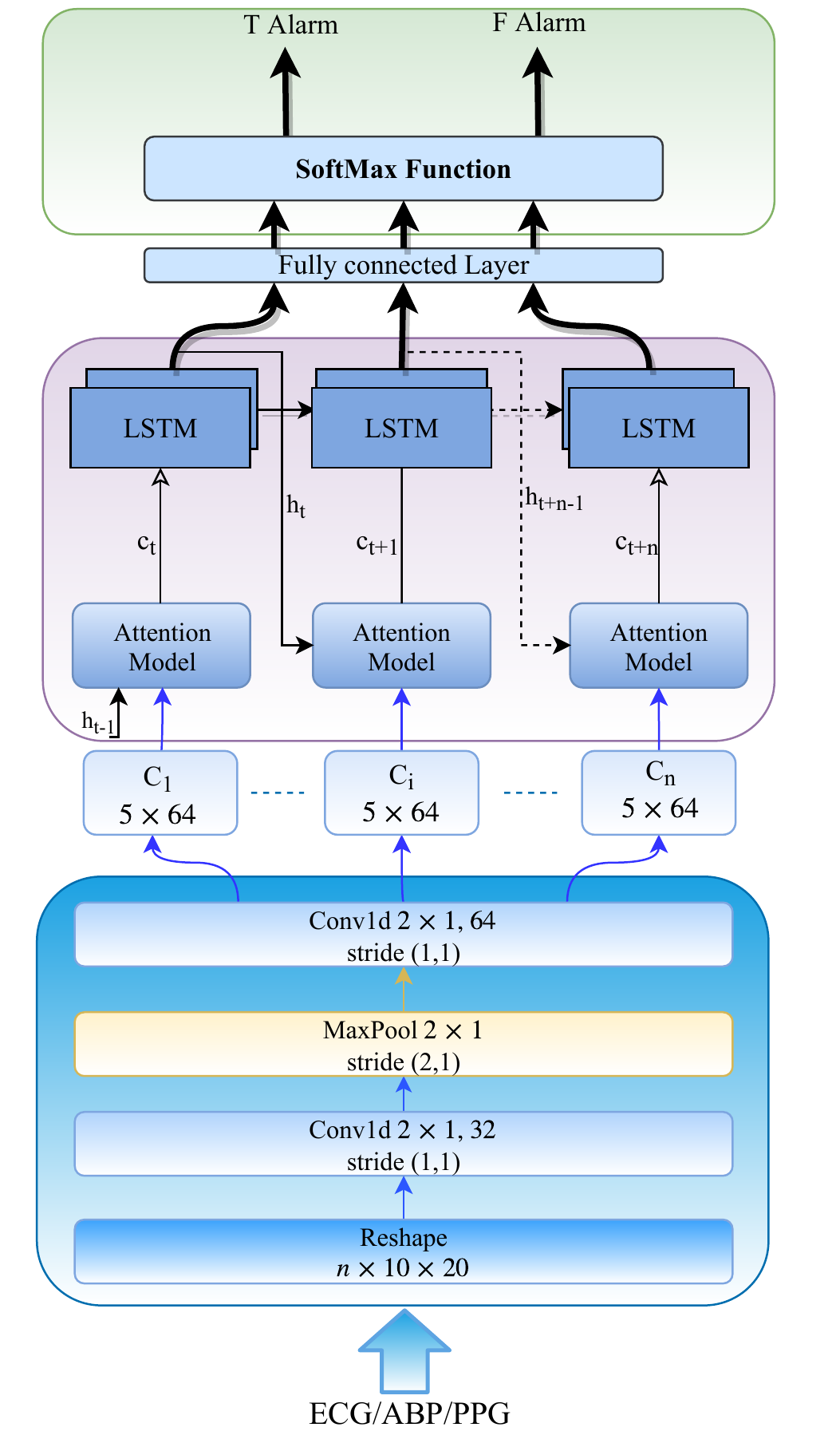}
  \caption{The network architecture of the proposed model for feature extraction.}
  \label{fig:model_ext}

\end{figure}

The following subsections describe the main parts of the automatic feature extraction network. We train a feature extraction network for each of the three input signals separately. Figure \ref{fig:model_ext} illustrates the proposed network architecture for automatic feature extraction. 
%First, we describe the main component of the network that we have trained as a feature extractor for each given signal. Then, we explain how we applied the extracted features to train a classifier to classify each three-input signal to a true or false alarm. Finally, we introduce a two-step training algorithm to train the proposed model.  
 
 \subsubsection{Convolutional neural network  (CNN)}
 \label{sec:cnn}
We employ two consecutive 1D convolutional layers with different sizes of filters and a max-pooling layer following the first convolutional layer. The first convolutional layer is composed of 32 filters with a kernel size of $2\times1$ and a stride 1, and a Rectified Linear Unit (ReLU) layer. The second convolutional layer with larger sizes of filters has 64 filters with a kernel size of $2\times1$ and a stride 1, and a ReLU layer. The max-pooling layer has a pooling region of size $2\times1$ with a stride size of $2\times1$. At each time step, a sequence of a segmented signal (e.g., ECG, ABP or PPG) with the size of n is fed to the CNN to extract features of interest. The second CNN layer generates $D$ feature maps of size $L\times1$ for each sample of the input signal, which is converted to L vectors of D-dimension as follows:
%The CNN part of the network is composed of two consecutive one-dimensional convolutional layers. The first layer has 32 filters with a kernel size of $2\times1$ and a stride 1 following by a Rectified Linear Unit (ReLU) layer. The second layer has 64 filters with a kernel size of $2\times1$ and a stride 1 following by an ReLU layer. The first layer is also followed a max pooling layer of pooling region of size $2\times1$ with a stride size of $2\times1$. At each time step, a sequence (with a size of n) of a segmented signal (e.g., ECG, ABP or PPG) is passed to the CNN to extract features. The last CNN layer produces $D$ feature maps of size $L\times1$ (e.g, here, we have $64$ feature maps with sizes of $5\times1$).
%Then, the feature maps are converted to $L$ vectors in which each vector has $D$ dimensions as follows: 
$$C_t = [C_{t,1},C_{t,2}, \ldots ,C_{t,L}], \qquad C_{t,i}\in \mathbb{R}^D.$$ 
%To provide better details, at each n time steps, there exists n $C_t$s (see Figure \ref{fig:model_ext}).
Here, we have $64$ feature maps with the sizes of $5\times1$ (see Figure \ref{fig:model_ext}).

 \subsubsection{Attention and Long Short-Term Memory (LSTM) units}
 \label{sec:att_lstm}
% \textcolor{magenta}{it seems like we have taken it for granted that using an attention layer will improve the performance. It would be better to have it as a hypothesis and evaluate the performance of the model with and without attention layer to show that it can effectively put a higher weight on windows contributing to the cardiac event. And even before that in the motivation section, add some papers from the medical community as an evidence that they believe there are some indicators in the physiological signals appearing before the actual conditions to justify the need for considering memory-based models like LSTM. In particular, the work that claimed that these indicators can appear in the signals as early as few hours before the cardiac event. Then, it would give us a good justification to say that since considering the entire length of the signal is not necessarily feasible, hence the soft attention can divide the signals to different partitions by putting a higher weight on the most important ones to save memory/computation and of course enhance the accuracy. You also need to add a discussion/justification on the length of $C_{t,i}$ and the impact of this selection on the performance. } 
We use an attention unit to learn the most effective parts of the input signal that are responsible to trigger a specific alarm. This unit assigns a probability value to each part of the signal to specify its importance in the prediction process (e.g., predicting true or false alarm). For instance, as depicted in Figure \ref{fig:model_ext}, the attention unit assigns a probability value to each vector extracted from the input segment by the CNN. Finally, an expected value of the most effective regions of the input segments is generated using the probability values provided by the attention units (represented by the feature vector, $C_t$).
 
  %Attention unit is used to learn the most relevant parts of an input data. It is done by assigning a probability value to each part. The probabilities indicate the importance of the parts in predicting a specific target (here, e.g., a true or false alarm). For example, as shown in Figure \ref{fig:model_ext}, the extracted vectors of each input segment (extracted by the CNN) are fed into the attention unit to calculate the probability values of vectors. Afterwards, the probability values are used to compute an expected value of the the most important regions of the input segment (represented by the feature vector, $C_t$).
 
 \begin{figure}[htb]
\centering
  \includegraphics[width=0.5\textwidth,height=1.0\textheight,keepaspectratio]{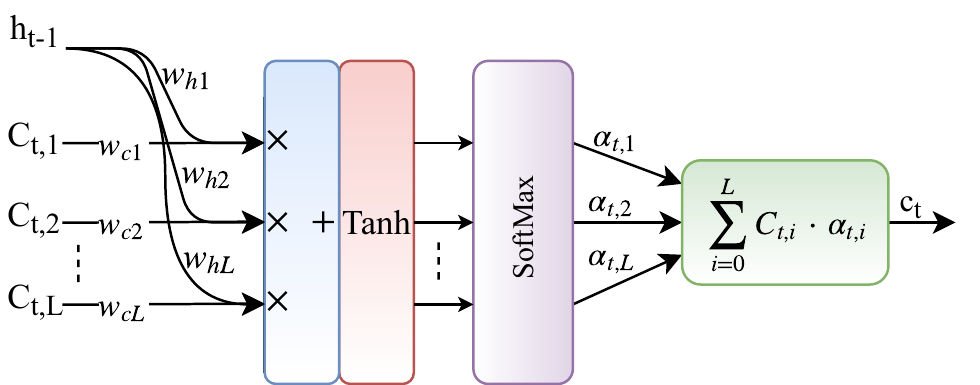}
  \caption{ A systematic diagram of the attention unit. The attention unit takes as input vertical
feature slices, $C_{t,i} \, i \in 1,2,\ldots ,{L},$ and the RNN previous hidden state, $h_{t-1}$. Then, it computes a linear weighted vector, $c_t$ that is a multiplication of each feature slice and its corresponding importance, $\alpha_{t,i}$.}
  \label{fig:attention_layer}
\end{figure}  

Figure \ref{fig:attention_layer} illustrates a systematic diagram of the attention unit utilized in our proposed model. The attention unit is fed by two inputs: (1) $L$ feature vectors, $C_{t,1},C_{t,2}, \ldots ,C_{t,L}$, where each $C_{t,i}$ represents a different part of the input segment, and (2) A hidden state $h_{t-1}$, which is the internal state of the RNN at the previous time step, $t-1$. Then, it calculates a vector, $c_t$ which is a weighted sum over feature slices, $C_{t,i}$. %linear weighted combination of the feature slices, $C_{t,i}$. 
With respect to the aforementioned assumptions, the attention mechanism can be formulated as:
 \begin{align}
  &\begin{aligned}
    \alpha_{t,i} & = f(\tanh({W_{h}} h_{t-1}+{W_{C}} C_{t,i}))       & i \in 1,2,\ldots ,{L},
  \end{aligned} \\
    &\begin{aligned}
   c_t & =\sum_{i=1}^{L} \alpha_{t,i} C_{t,i},
  \end{aligned}
\end{align}
% \colorbox{yellow}{define all these parameters}
In the above equations, $\alpha_{t,i}$ is the importance of  part $i$ of the input segment. $f(.)$ is a softmax function that processes a vector of L real numbers as input, and normalizes them into probability values. At first, a vector consisted of a weighted sum over $C_{t,i}$ and $h_{t-1}$ values is created and passed to the $\tanh$ function. Then, the softmax function normalizes the ${L}$ values of the input vector and creates $\alpha_{t,i}$. 
 %where $\alpha_{t,i}$ is the importance of the part $i$ of the input segment. $f(.)$ is a softmax function that gets a vector of L real numbers as input, and normalize them into probability numbers. At each time step $t$, the attention unit calculates a vector of a composition of the $C_{t,i}$ values and $h_{t-1}$ followed by a $\tanh$ function. Then, the vector is fed into a softmax function to generate $\alpha_{t,i}$ over ${L}$ parts. 
 In other words, each $\alpha_{t,i}$ is considered as the amount of importance of the corresponding vector $C_{t,i}$ among ${L}$ vectors in the input segment. Finally, the attention unit calculates $c_t$, a weighted sum of all vectors $C_{t,i}$ with respect to $\alpha_{t,i}$s. Following the above process, the model attempts to learn to put more emphasis on the important regions of the input segment with higher probabilities that make to trigger an alarm (e.g., a false or true alarm) in ICUs.
%Note that the soft attention model is fully differentiable which allows training the systems in an end-to-end manner.
 
In order to extract temporal information and capture long-range of dependencies between segments of the input signal, we employ a stack of two long short-term memory (LSTM) units with sizes of 256. The LSTM units are following the attention units and take $c_{t+i}$ values produced by the attention units and the previous hidden states of the LSTM units as inputs to generate the next hidden states. In other words, the LSTM unit takes $c_{t}$ , the output of attention unit at time $t$, and $h_{t-1}$, previous hidden state, to return the next hidden state $h_{t}$. The new hidden states are fed to the attention units to produce the value of $h_{t}$ at the next step and also the fully-connected layer with a size of 256 (see Figure \ref{fig:model_ext}).

%In addition, we applied long short-term memory (LSTM) units to extract temporal information and dependencies between segments of each input signal. The attention units are followed  by a stack of two LSTM units with sizes of 256. As illustrated in Figure \ref{fig:model_ext}, the LSTM units sequentially takes as input $c_{t+i}$ values provided by attention modules and the previous hidden states of the LSTM units, and output the next hidden states. In other words, at each time step $t$, each LSTM unit takes as input the output of attention unit, $c_{t}$ and a previous hidden state $h_{t-1}$, and computes the next hidden state $h_{t}$. The new hidden state is used by the attention unit to generate the value of $c_{t+1}$ at the next time step, and also utilized as input of a fully-connected layer with size of 256 units. 
 
\subsubsection{Classification layer}
\label{sec:class_layer}
This layer specifies the label of the input signal (i.e., true or false alarm) and consists of a fully-connected layer followed by a softmax layer. The softmax layer assigns probabilities that the given input belongs to each of the class labels (i.e., true or false alarm classes). Note that this layer is removed while the model depicted in Figure \ref{fig:model_ext} is used as a feature extractor in the network illustrated in Figure \ref{fig:final-model}.
\subsection{Loss calculation}
\label{sec:loss}
An important caveat in the false alarm reduction research is the class imbalance problem, meaning that the number of true alarms is much less than the false alarms. This problem causes to drop the performance of the applied method for the minor class. To tackle this problem, we utilize a combination of two loss functions of mean false error (MFE) and mean squared false error (MSFE) \cite{mousavi2019sleepeegnet,wang2016training} instead of the commonly used Mean Squared Error (MSE) in deep learning algorithms. These loss functions calculate the training error without considering the membership of the misclassified sample to the major or minor class. In other words, the MFE and MSFE methods capture the training error of the classes equally as opposed to the MSE method that is biased to the major class in a imbalanced dataset. The loss functions can be defined as follows:
%False alarm reduction problem suffers from the problem of class imbalance where the number of false alarms are much greater than true alarms (see Table \ref{tab:StatDataset}). To deal with this issue, we apply some loss functions (rather  than  commonly  used  mean  squared  error  (MSE)  in deep learning algorithms.) to alleviate the effect of this imbalanced class problem on the model training \cite{mousavi2019sleepeegnet,wang2016training}. These loss functions are called mean false error (MFE) and mean squared false error (MSFE). They treat the captured error of each misclassified sample equally regardless of being in the majority or minority class. That is to say, the  MFE  and  MSFE  attempt  to  treat  the  errors  of  all  classes equally, unlike the mean squared error (MSE) that is biased to the  majority  classes in  the  imbalanced  dataset. Therefore, the loss functions can be defined as:

% it is computed by  averaging  the  squared  difference  between predictions labels  and given labels. Therefore, it makes the contribution of  the  majority  classes  be  much  more  in  comparison  with the  minorities  classes  in  the  imbalanced  dataset.

\begin{align}
 &\begin{aligned} 
l(g_i)= \frac{1}{G_i} \sum_{i=j}^{G_i} (y_j-\hat{y_j})^2,
  \end{aligned}\\
 &\begin{aligned} 
l_{MFE}= \sum_{i=1}^{N} l(g_i),
  \end{aligned}\\
 &\begin{aligned} 
l_{MSFE}= \sum_{i=1}^{N} l(g_i)^2,
  \end{aligned}
\end{align}

In the above equations, $g_i$ is the class label (e.g., true or false alarm), $G_i$ is the number of samples in the class $g_i$, $N$ is the number of available classes (in this study, we have two classes), and $l(g_i)$ is the error calculated over the class $g_i$.
%where $g_i$ is the class label (e.g., T or F alarm), $G_i$ is the number of the samples in class $g_i$, $N$ is the number available classes (i.e., two classes in this context), and $l(g_i)$ is the calculated error for the class  $g_i$. 

\subsection{Training algorithm}
% \textcolor{red}{select the model (fold) with the best F1-score!}
In order to effectively train the proposed model via back-propagation algorithm, we present a two-step training algorithm as illustrated in \ref{alg:training_alg}. Step 1 (lines 1-9) involves extracting the features of interest for a specific input signal (i.e., for each of ECG, ABP, and PPG signals, separately). Then, pre-trained networks are used as feature extractors for their corresponding models including ECG, ABP, and PPG. In this step, in order to apply the pre-trained networks as feature extractors, only the output of the fully-connected layer in the classification layer is utilized to represent the given signal and the softmax layer is discarded (i.e., line 8). In step 2 (lines 10-16), the classification task is accomplished using the three signals as shown in Figure \ref{fig:final-model}. It must be pointed out that the three pre-trained networks are frozen during training process and the second part of the model is trained to generate a label. Also, training the models in both steps are performed with the same hyper-parameters. 

 %We develop a two-step training algorithm as shown in Algorithm \ref{alg:training_alg}. At the first step of the algorithm (lines 1-9), the model shown in Figure \ref{fig:model_ext} is trained for each specific signal including ECG, ABP and PPG, separately. Then, each trained model is utilized as a feature extractor for its corresponding modal (i.e., ECG, ABP or PPG) in the next step of the algorithm. It is important to note that to use the pre-trained models as feature extractors, the softmax layers of the models (after feature learning is done) are discarded, and just the output of the fully-connected layers are considered as the representation of the given signal. 
 %At the second step of the algorithm (lines 10-16), the model shown in Figure \ref{fig:final-model} is trained to generate a label regarding the three give signals. During training of the model, the three pre-trained networks are frozen and only the rest of the model is trained. Note, the same hyper-parameters are used for training the models in both steps.
 
 \begin{algorithm} \caption{Two-step training algorithm for the proposed model}
\begin{algorithmic}[1]
% \Require  hyper-parameters, data
\Input{hyper-parameters, data}
\Output{f\_model}
% \STATE  Output: final model
% \State Initilize $V(s) = 0$, for all $s \in \mathcal S^+$
\Statex \textbf{\textit{Step 1:}}
\ForEach {$modal$ \textbf{in} $[ECG, ABP, PPG]$}
\State Initialize $NET[modal]$ with random weights
\For {$i=1$ \textbf{to} $n\_epochs$}
\ForEach {$batch$ \textbf{in} $batch\_data(data,modal)$}
\State \vspace{-.6cm} \begin{align*} 
 \qquad \qquad \quad  NET[modal] \gets & train\_network
 \\ &  (NET[modal],batch), \\ & as ~ shown ~ in ~ Figure~ \ref{fig:model_ext}\end{align*}
\vspace{-.6cm}
\EndFor
\EndFor
\State $NET[modal]  \gets r\_softmax\_layer(NET[modal])$
\EndFor

\Statex \textbf{\textit{Step 2:}}
\State Initialize $f\_model$ with random weights
\For {$i=1$ \textbf{to} $n\_epochs$}
\ForEach {$batch$ \textbf{in} $batch\_data(data)$}
\State \vspace{-.6cm} \begin{align*}
\qquad \quad f\_model  \gets & train\_model(f\_model,NET[ECG],\\
&NET[ABP], NET[PPG],batch),\\ & as ~ shown ~ in ~ Figure~ \ref{fig:final-model}
\end{align*}
\vspace{-.6cm}
\Comment{Learning for $NET[.]$ is frozen.}
\EndFor
\EndFor
\State \textbf{return} $f\_model$
\end{algorithmic}
 \label{alg:training_alg}
\end{algorithm}

\section{Dataset}
\label{sec:Dataset}
\begin{figure}[htb]
\centering
  \includegraphics[height=1.\textheight,width=1.\linewidth,keepaspectratio]{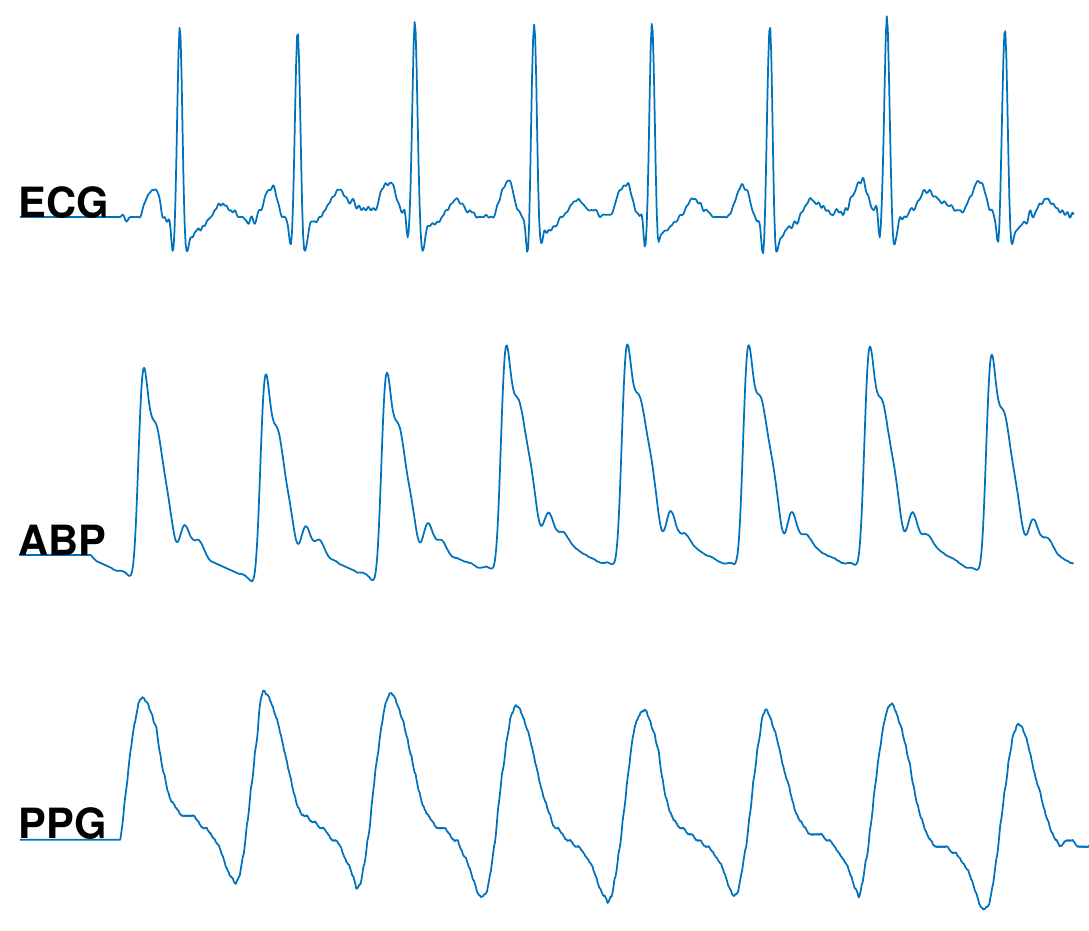}
  \caption{Illustration of an electrocardiogram (ECG), an arterial blood pressure (ABP) and a photoplethysmogram (PPG) signal.
      }
  \label{fig:ecg_abp_ppg}
\end{figure}

\begin{figure}[bt]
\centering
 \includegraphics[height=1.\textheight,width=1.\linewidth,keepaspectratio]{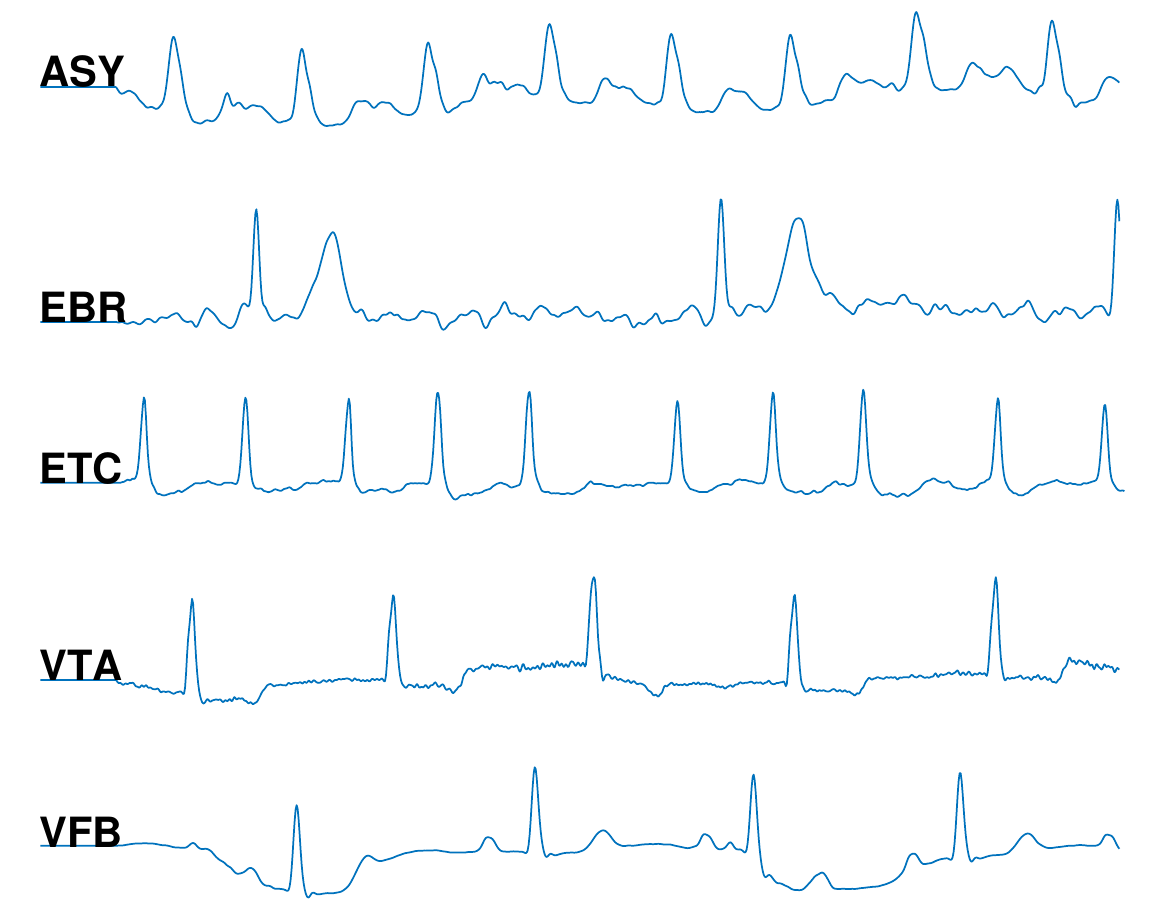}
  \caption{Five common critical alarm types in the intensive care units as used in the PhysioNet/Computing in Cardiology Challenge 2015 \cite{PhysioNetFL}.}
%   \textcolor{magenta}{It would be good to add a table here to have the definition and the samples in the same place}
  \label{fig:ASY_EBR_ETC_VTA_VFB}
\end{figure}

We applied the publicly available alarm database for ICUs provided by PhysioNet computing in cardiology challenge 2015 \cite{PhysioNetFL,clifford2015physionet}. It includes five types of life-threatening arrhythmia alarms: Asystole (ASY), Extreme Bradycardia (EBR), Extreme Tachycardia (ETC), Ventricular Tachycardia (VTA), and Ventricular Flutter/Fibrillation (VFB). The definition and visualization of each alarm are presented in Table \ref{tab:alarm-def} and in Figure \ref{fig:ASY_EBR_ETC_VTA_VFB}, respectively. The training set includes 750 recordings and the test set includes 500 recordings. The test set has not been publicly available yet, therefore we use the training set for both test and training purposes. Each is recording composed of two ECG leads and one or more pulsatile waveforms (i.e., the photoplethysmogram (PPG) and/or arterial blood pressure (ABP) waveform).  Figure \ref{fig:ecg_abp_ppg} shows a sample of each type of the ECG, ABP and PPG signals.
The signals were re-sampled to a resolution of 12 bit and frequency of 250 Hz and filtered by a finite impulse response (FIR) bandpass [0.05 to 40 Hz] and mains notch filters for denoising. The alarms were labeled with a team of expert to either 'true' or 'false'. Table \ref{tab:StatDataset} shows the statistics of the numbers of true and false alarms of each arrhythmia type in the training set. 
\begin{table*}[htb]  
\caption{Alarms definition}
\centering{
\label{tab:alarm-def}
	\resizebox{1.\linewidth}{!}{  %fit to windows command 
\begin{tabular}{lc}
\toprule
\textbf{Alarm Type} & \textbf{Definition}\\ 
\midrule
\texttt{Asystole(ASY)} &There might not be heartbeats for more than 4s in the signal \\
\texttt{Extreme Bradycardia (EBR)} &The heart rate is less than 40 beats per minute (bpm)  \\ 
\texttt{Extreme Tachycardia (ETC)} & The heart rate would be greater than 140 bpm for 17 consecutive beats  \\ 
\texttt{Ventricular Tachycardia (VTA)} & A sequence of five or more ventricular beats with the heart rate greater than 100 bpm in the signal  \\ 
\texttt{Ventricular Flutter/Fibrillation (VFB)} & A rapid Fibrillatory, flutter, or oscillatory waveform for at least 4 seconds in the signal  \\ 
\bottomrule  
HR: Heart rate \\ 
\end{tabular}}
}
\end{table*}

\begin{table*}[hbt]
\centering{
    \caption{The statistics of the numbers of true and false alarms of each arrhythmia type.}
    \label{tab:StatDataset} 
    \resizebox{0.8\linewidth}{!}{ 
    \begin{tabular}{l|c|c|r} % <-- Alignments: 1st column left, 2nd middle and 3rd right, with vertical lines in between
      \textbf{Alarm} & \textbf{\# of patients} & \textbf{\# of false alarms} & \textbf{\# of true alarms}\\
       \hline
      Asystole (ASY) & 122 & 100 & 22\\
      Extreme Bradycardia (EBR) & 89 & 43 & 46\\
      Extreme Tachycardia (ETC) & 140 & 9 & 131\\
      Ventricular Tachycardia (VTA) & 341 & 252 & 89\\
      Ventricular Flutter/Fibrillation (VFB) & 58 & 52 & 6\\
      \hline
      Total & 750 & 456 & 294\\
    \end{tabular}
    }
}
\end{table*}

% Five types of life-threatening arrhythmia alarms, can be defined as follows. \textit{Asystole} (ASY), which means there might not be heartbeats for more than 4s in the signal, \textit{Extreme Bradycardia} (EBR), which means the heart rate is less than 40 beats per minute (bpm), \textit{Extreme Tachycardia} (ETC), which means the heart rate would be greater than 140 bpm for 17 consecutive beats. \textit{Ventricular Tachycardia}(VTA), which means there is a sequence of five or more ventricular beats with heart rates greater than 100 bpm in the signal, and finally, \textit{Ventricular Flutter/Fibrillation} (VFB), which means there exists a rapid Fibrillatory, flutter, or oscillatory waveform for at least 4 seconds in the signal (see Figure \ref{fig:ASY_EBR_ETC_VTA_VFB}). 

%  One of our aims in this study is to develop more accurate method to the identification of these five false arrhythmia alarms (i.e., the problem of high false alarm rates) using processing given physiological signals. 
 
\section{Experimental Results}
\label{sec:expres}

\subsection{Experimental Design}
The performance of the proposed model was evaluated using the PhysioNet challenge-2015 dataset. Since multi-modal prediction is based on the three signals of ECG, ABP and PPG, only 220 samples out of 750 recordings that include all these signals are used and for the single-modal method all samples are utilized. The PhysioNet challenge 2015 \cite{clifford2015physionet} have considered two main events: (i) real-time setting in which the information before the alarm onset can be used, and (ii) retrospective setting in which up to 30 seconds of data after the alarm can be used. In this study, we focus on the real-time setting where only information prior to occurring the alarm is used. As mentioned above, using all signals in the learning process makes the model take benefit of all available information and extract the correlation between different models. We used $k$-fold cross-validation approach to train and test the proposed model with a $k$ size of 10 unless explicitly stated otherwise. Indeed, we divided the dataset into k= 10 folds. Then, for each fold of the 10 folds, one fold is used for evaluating the model and the remaining 9 folds are used to train the model. In the end, all evaluation results were concatenated.

Both whole model and the three networks (ECG, ABP and PPG Nets) were trained with a maximum of 100 epochs and a mini-batch size of $10$.
 The RMSProp optimizer was applied to minimize the $l_{MFE}$ loss with a learning rate parameter of $\alpha = 0.001$. 
 Two different regularization techniques were used to prevent the overfitting problem. First, the dropout layer with the probability of dropping of 0.5 (as shown in Figure \ref{fig:final-model}). At every learning iteration, the dropout function chooses the some nodes randomly and deletes them along with their connections. Second, an additional $L_2$ regularization term with $\beta = 0.001$ was added to the loss function. This kind of regularization tries to punish the model parameters with large values.  As a result, it prevents an  unstable learning (i.e., the exploding gradient problem). Python programming language along with Google Tensorflow deep learning library were used to implement our model. Furthermore, a machine with 8 CPUs (Intel(R) Xeon(R) CPU @ 3.60 GHz), 32 GB memory and Ubuntu 16.04 was utilized to run the k-fold cross validation.
 The training time for each epoch was 98 seconds on average and the testing time for each batch of 20 EEG epochs was approximately 0.102 seconds.
 
Different metrics were considered to assess the performance of the proposed model. These metrics include accuracy (ACC), sensitivity (SEN), specificity (SPE), precision (PRE), F1-score, and area under the ROC curve (AUC). We also report the PhysioNet Challenge 2015 score for our proposed method. It is defined as $score = (TP + TN)/(TP +TN + FP + 5 \times FN)$, where TP is true positives, FP is false positives, FN is false negatives, and TN is true negatives. All results are reported as an average over k-folds, where k can set to 5 and 10). 

\subsection{Results and Discussion}
The results in Table \ref{tab:compare} represent the alarm classification success for our proposed method against other methods in the literature while three signals (i.e., ECG II, ABP and PPG) are considered. It can been seen from the table that our model significantly outperforms other methods. We also experimented our single-modal (using just one single lead) approach to bold how outcome might be different. Table \ref{tab:compare} demonstrates using the multi-modal approach absolutely leads in better performance results compared to the single-modal one. 

The results provided in Table \ref{tab:compare} are for 220 samples of dataset with three available signals, aggregating all alarm types. We also evaluated our model with samples with just Ventricular Tachycardia alarm type. There were two main reasons that we selected this alarm type, (1) the number of samples for other life-threatening arrhythmia alarm types were too small, Asystole (34: 4 true and 30 false alarms), Extreme Bradycardia (30: 21 false and 9 true alarms), Extreme Tachycardia (15: 14 false and 1 true alarms), Ventricular-Flutter/Fibrillation (17: 12 false and 5 true alarms), and Ventricular Tachycardia (124: 106 false and 18 true alarms), (2) the Ventricular Tachycardia alarms are more difficult than other alarm types to detect \cite{clifford2015physionet}. Table \ref{tab:compare_vt} shows the performance of our proposed model for Ventricular Tachycardia alarm type using a single-lead signal and multi-lead signals. Our method achieves remarkable results for both the multi-modal and the single-modal (ECG II) approaches, a sensitivity and a specificity of 93.75\% and 93.92\% for the single-modal technique, and a sensitivity and a specificity of 93.75\% and 95.49\% for the multi-modal technique. As shown in the table, our method outweighs the other method significantly. It also can be see that using all available signals performs better compared to the single-lead signal. 

We also investigated how our model behaves for all alarm types using single-lead ECG waveforms. Table \ref{tab:compare_alarms} compares the performance (in terms of true positive rate (TPR or also called the sensitivity) true negative rate (TNR or also called specificity) and AUC) of various algorithms using different signals. As can be seen in Table \ref{tab:compare_alarms}, the proposed method performs better than the methods proposed by Lehman et al. \cite{lehman2018representation} and Li et al. \cite{li2017false} on Ventricular Tachycardia (VTA) alarm. 
% (the only available false alarm reduction results in the literature).
Furthermore, our method using single-lead ECG (ECG II) detects Extreme Bradycardia (EBR),  Extreme Tachycardia (ETC) and  Ventricular-Flutter/Fibrillation (VFB) alarms significaly better than other methods using two-lead ECG (Lehman et al. \cite{lehman2018representation}) and all available signals, including ECG II, ECG V, ABP and PPG (Ansari et al. \cite{ansari2016suppression} and Gajowniczek et al. \cite{gajowniczek2019reducing}). Moreover, as shown in Table \ref{tab:compare_alarms}, our proposed single-modal method leads to comparable results (in some cases, even better outcomes) for detecting Asystole (ASY) and Ventricular Tachycardia (VTA) arrhythmical alarm types compared to other listed algorithms that have utilized more than one signal. In addition, we note that here our remarkable results were obtained using a single-lead ECG (ECG II), however having more than one modal would leads to a improvement in performance results.

Furthermore, Table \ref{tab:per_alarms} reports the evaluation results of our single-modal proposed method with various metrics, including the  challenge score provided by the PhysioNet Challenge 2015, using just the ECG II signal. This table can be used as a reference to compare future work.

% "...It is multi-stage approach unlike our approach..."

\begin{table*} [htb] 
\caption{Comparison of performance of the proposed model against other algorithms on  the PhysioNet challenge-2015 dataset.}
 \centering{
\label{tab:compare}
\resizebox{1.\linewidth}{!}{  %fit to windows command 
\begin{tabular}{lcccccccccc}
% \toprule
\toprule[\heavyrulewidth]\toprule[\heavyrulewidth]
\textbf{} & \textbf{} &  \textbf{} & \textbf{} &\multicolumn{6}{c}{\textbf{Best Performance (\%)}} \\
\cmidrule(lr){5-10}
\textbf{Method} &   \textbf{Signal}& \textbf{\# of samples}&  \textbf{CV} &  {$SEN$} & {$SPE$} &{$PRE$} &{$ F1-score$}&{$AUC$}& {$ACC$}\\
\midrule
\textbf{Multi-modal method}&All&220&10-fold CV  &\textbf{\underline{93.88}} &\textbf{\underline{92.05}} &\textbf{\underline{79.31}}  &\textbf{\underline{85.98}} &\textbf{\underline{92.99}} &\textbf{\underline{92.50}} \\
Zaeri-Amirani et al. \cite{zaeri2018feature}&All&220&10-fold CV  & 73 &75&-  &- &81 &77\\
Afghah et al. \cite{afghah2018game}& All&220&10-fold CV   & 80 & 71 &-  &- &74.32 & 77.6 \\
\hline
\textbf{Single-modal method}& ECG II&220&10-fold CV  &73.33 &87.74 &63.46  &68.04 &80.53 &84.50 \\
\textbf{Single-modal method} & ABP&220&10-fold CV  &78.72 &65.35 &41.11  &54 &72.04 &68.50 \\
\textbf{Single-modal method} & PPG&220&10-fold CV  &87.50 &63.15 &42.96  &57.53 &75.32 &69 \\

 \bottomrule  
\multicolumn{3}{c}{All: ECG II, ABP, PPG; CV: Cross Validation}
%  \multicolumn{8}{c}{ CV: Cross Validation}%; Tr: Training; Te: Test}
\end{tabular} }
}
\end{table*}

% \textcolor{red}{Remember to put loss and accuracy curves!}
% \textcolor{red}{mention different batch-sizes and the reason!}

\begin{table*} [htb] 
\caption{Comparison of performance of the proposed model against other algorithms for alarm type of Ventricular Tachycardia arrhythmia on the PhysioNet challenge-2015 dataset.}
 \centering{
\label{tab:compare_vt}
\resizebox{1.\linewidth}{!}{  %fit to windows command 
\begin{tabular}{lcccccccccc}
% \toprule
\toprule[\heavyrulewidth]\toprule[\heavyrulewidth]
\textbf{} & \textbf{} &  \textbf{} & \textbf{} &\multicolumn{6}{c}{\textbf{Best Performance (\%)}} \\
\cmidrule(lr){5-10}
\textbf{Method} &   \textbf{Signal}& \textbf{\# of samples}&  \textbf{CV} &  {$SEN$} & {$SPE$} &{$PRE$} &{$ F1-score$}&{$AUC$}& {$ACC$}\\
\midrule
\textbf{Multi-modal method}&All&124/220&10-fold CV  &\textbf{\underline{93.75}} &\textbf{\underline{95.49}} &\textbf{\underline{85.41}}  &\textbf{\underline{86.67}} &\textbf{\underline{94.61}} &\textbf{\underline{95}} \\
Afghah et al. \cite{afghah2018game}& All&124/220&10-fold CV  &86 &- &73  &- &- &85.48 \\
\hline
\textbf{Single-modal method}& ECG II&124/220&10-fold CV  &93.75 &93.92 &79.16  &84.58 &93.84&93.75 \\
\textbf{Single-modal method} & ABP&124/220&10-fold CV  &81.25 &75.68 &41.95  &69.76 &78.46 &76.67\\
\textbf{Single-modal method} & PPG&124/220&10-fold CV  &100 &50 &33.33 &50 &75 &60 \\

% \hline
% \textbf{Single-modal method}&ECG II&341/750&10-fold CV  &{\underline{89.86}} &{\underline{88.56}} &{\underline{72.94}}  &{\underline{80.52}} &{\underline{89.21}} &{\underline{88.89}} \\
% \textbf{Single-modal method}& ABP&341/750&10-fold CV  &- &- &-  &- &-&- \\
% \textbf{Single-modal method}&PPG&341/750&10-fold CV  &- &- &-  &- &-&- \\
% Gajowniczek et al. \cite{gajowniczek2019reducing} &All&341/750&10-fold CV  &67.8 &88.9 &-  &- &87.00&- \\

% Li et al. \cite{li2017false}& ECG II&341/750&0.67/0.33 &76.70 &59.80 &39.80  &52.40 &- &- \\
 \bottomrule  
\multicolumn{3}{c}{All: ECG II, ABP, PPG; CV: Cross Validation}

\end{tabular} }
}
\end{table*}

\begin{table*} [htb] 
\caption{Comparison of performance of the proposed model against other algorithms for all alarm types on the PhysioNet challenge-2015 dataset.}
 \centering{
\label{tab:compare_alarms}
\resizebox{1.\linewidth}{!}{  %fit to windows command 
\begin{tabular}{l*{18}{c} }
% \toprule
\toprule[\heavyrulewidth]\toprule[\heavyrulewidth]
& & &\multicolumn{3}{c} {ASY} & \multicolumn{3}{c} {EBR} & \multicolumn{3}{c} {ETC} &  \multicolumn{3}{c} {VTA} &  \multicolumn{3}{c} {VFB} \\
\cmidrule(lr){4-6} \cmidrule(lr){7-9} \cmidrule(lr){10-12} \cmidrule(lr){13-15} \cmidrule(lr){16-18}
\textbf{Method} & Signal &CV&TPR &TNR&AUC & TPR& TNR&AUC &TPR& TNR&AUC& TPR& TNR&AUC &TPR &TNR&AUC\\
\midrule
% \textbf{Single-modal method} & - &- & -& - &-& -& -& - &-&-&-&-&-&- &-\\
\textbf{Single-modal method} &ECG II&5-fold
&\textbf{\underline{96.67}} &\textbf{\underline{82.16}} & \textbf{\underline{89.41}}& \textbf{\underline{97.78}} &\textbf{\underline{94.85}}& \textbf{\underline{96.31}}& \textbf{\underline{100}}& \textbf{\underline{100}}&\textbf{\underline{100}}&\textbf{\underline{90.71}}&\textbf{\underline{88.30}}&\textbf{\underline{89.51}}&\textbf{\underline{100}}&\textbf{\underline{97.22}}&\textbf{\underline{98.61}}\\

Lehman et al. \cite{lehman2018representation}&ECG II*& 10-fold &-&- & -& - &-&-& -&-& -& -&-&87 &-&-&-\\

Li et al. \cite{li2017false}&ECG II&0.67/0.33 &- &-&- & -& - &-&-& -&-& 76.70 &59.80&-&- &-&-\\
\hline
Lehman et al. \cite{lehman2018representation}&ECG II/V*& 10-fold &-&- & -& - &-&-& -&-&-& 89& 86&91&-&-&-\\
Ansari et al. \cite{ansari2016suppression}  &All&5-fold &84.97 &89.21&- & 90.49& 90.05 &-&96.55& 97.80&-& 96.63& 95.47&-&92.40 &61.64&-\\

Gajowniczek et al. \cite{gajowniczek2019reducing}  &All&10-fold &85 &90&95 & 84.5& 91 &93.3&99.2& 77.8&99& 67.8& 88.9&87&83.3 &94.2&95\\

\bottomrule  
\multicolumn{16}{c}{All: ECG II, ECG V, ABP, and PPG; CV: Cross Validation; *: 1250 records (750 train, 500 hidden test of Physionet), in which 562 records contains VTA alarms} 

\end{tabular} }
}
\end{table*}

\begin{table} [htb] 
\caption{Performance of the proposed model for all alarm types on the PhysioNet challenge-2015 dataset, considering just a single lead (ECG II).}
 \centering{
\label{tab:per_alarms}
\resizebox{1.\linewidth}{!}{  %fit to windows command 
\begin{tabular}{l*{8}{c} }
% \toprule
\toprule[\heavyrulewidth]\toprule[\heavyrulewidth]
& \multicolumn{6}{c} {Best Performance (\%)} \\
\cmidrule(lr){2-8} 
\textbf{Alarm}  &SEN& SPE &PRE &F1-score &AUC &ACC& Score\\
\midrule
\textbf{ASY} & 96.67 &82.17 & 57.33& 69.21 &89.41& 84.20&81.20\\
\textbf{EBR} & 97.78 &94.85 & 93.76&95.56& 96.31 &96& 92.35\\
\textbf{ETC} & 100 &100 & 100& 100 &100& 100&100\\
\textbf{VTA} & 90.71 &88.30 & 74.88& 81.41 &89.51& 88.89&81.55\\
\textbf{VFB} & 100 &97.22 & 87.50& 91.67 &98.61& 97.50&97.50\\

\bottomrule  
\multicolumn{6}{c}{Score: PhysioNet/CinC Challenge 2015 Score}

\end{tabular} }
}
\end{table}

\section{Conclusion}
\label{sec:con}

False arrhythmia alarm reduction in ICUs is a challenging classification problem because of the presence of different sources of noise and artifacts in the data (i.e., the collected signals) as well as a large number of false alarms that results in the class imbalance problem. In this study, we proposed a deep learning-based network composed of the CNN layers, attention mechanism, and LSTM units to reduce false alarm arrhythmia in ICUs. We also utilized a new loss function to alleviate the effect of the class imbalance problem while training the model. Our proposed approach utilized a two-step training algorithm that trains the model for each modal (i.e., ECG, ABP, and PPG) to efficiently extract features, and then uses the combined features of each modal to classify the three-input signal to a true or false alarm (i.e., in a multi-modal way). Our proposed multi- and single-modal approaches demonstrated high performance for the suppression of false alarms without disregarding the true alarms compared to the existing algorithms in the literature.

% or


% \appendices

% % you can choose not to have a title for an appendix
% % if you want by leaving the argument blank
% \section{}

% \section*{Acknowledgment}
% This material is based upon work supported by the National Science Foundation under Grant Number 1657260. Research reported in this publication was supported by the National Institute On Minority Health And Health Disparities of the National Institutes of Health under Award Number U54MD012388.

\ifCLASSOPTIONcaptionsoff
  \newpage
\fi

\bibliographystyle{IEEEtran}
\bibliography{bare_jrnl}
\end{document}